\documentclass[9pt,twocolumn,twoside]{osajnl}

\journal{ol} 

\setboolean{shortarticle}{true}

\usepackage[justification=justified]{caption}
\usepackage{mathrsfs}

\title{Experimental generation of a flat-top beam profile in a stable ring cavity}

\author[1]{A. López-Vázquez}
\author[1]{Y. M. Torres}
\author[1]{M.S. Billión}
\author[1]{W. M. Pimenta}
\author[2]{J.A. Franco-Villafañe}
\author[1,*]{E. Gomez}

\affil[1]{Instituto de Física, Universidad Autónoma de San Luis Potosí, San Luis Potosí 78290, México}
\affil[2]{CONACYT - Instituto de Física, Universidad Autónoma de San Luis Potosí, San Luis Potosí 78290, México}

\affil[*]{Corresponding author: egomez@ifisica.uaslp.mx}

\begin{abstract}
We create a spatially homogeneous field inside of a ring cavity by combining two transverse modes generated by a single laser through modulation. The interference term between the two modes averages out because of the frequency difference between them, eliminating the need for interferometric control of their relative phase. The use of a ring cavity allows for a large waist for the flat-top profile, big enough to cover the atoms in an atomic trap. The cavity is mechanically and thermally isolated, and the laser light is locked to the cavity using the Pound-Drever-Hall technique. The flat-top profile technique reported here fulfill the vanishing curvature criterion at the center of the profile. 
\end{abstract}

\setboolean{displaycopyright}{true}

\begin{document}

\maketitle

\section{Introduction}
Optical cavities have been extensively used in recent years. They offer certain advantages over free propagating beams in atomic manipulation (preparation, cooling, feedback, and monitoring), inducing collective behavior and producing novel quantum phases \cite{ritsch13}, achieving sub-recoil cooling \cite{wolke12}, among others. In contrast to a Fabry-Perot cavity where a standing field drives the dynamics, the stable ring cavity takes advantage of traveling field modes that display differences in, for example, the collective modes that can be excited \cite{chen10} or the way fermions scatter light \cite{meiser05}. The light beams in a ring cavity can have a bigger waist, which helps to reduce heating in the mirrors for high power laser applications \cite{carstens13}. There are laser cooling schemes designed for ring cavities \cite{elsasser03}, although cavity cooling has also been demonstrated for Fabry-Perot cavities \cite{wolke12}. For matter-wave interferometry, the use of a ring cavity provides some advantages \cite{riou17}. On one hand, large momentum transfer can be achieved exploiting the power enhancement of the cavity. On the other hand, the wavefront distortions are reduced by the spatial filtering on the cavity modes. Recently, a malleable potential has been produced in a bow-tie cavity to split a Bose-Einstein condensate (BEC) by exciting simultaneously multiple transverse modes \cite{naik18}.\\
It is of particular interest the generation of flat-top beams.
They give a homogeneous intensity profile that increases the contrast in atomic interferometers \cite{mielec18}. Significantly lower density is observed in a BEC when a uniform potential is used instead of a harmonic one \cite{gaunt13}, and it has been proposed that a BEC in a homogeneous potential can be used as a gravitational wave detector \cite{sabin14}. Flat-top profiles have been produced in many ways, including appropriate amplitude and/or phase elements \cite{dickey}, spatial light modulators \cite{ngcobo13,liang09,bhebhe18} or optical feedback in a microchip laser \cite{naidoo16}. Some experimental demonstrations for intra-cavity flat-top beams are achieved by placing optical elements inside the cavity like a gain medium besides diffractive elements \cite{naidoo12}, or by using non-conventional mirror surfaces that can produce a flat-top at a particular position inside the cavity \cite{litvin09}. To the best of the authors' knowledge, this work is the first experimental realization of a flat-top profile inside an empty ring cavity with conventional mirrors. Compared to other methods, using the ring cavity allows for a high-quality factor and a very clean mode. The techniques developed in this paper are not limited to cold gases in cavities, the same ideas could be extended to other highly sensitive experiments such as gravitational wave detection to reduce thermal noise fluctuations \cite{vinet05}.\\
In this paper, we experimentally demonstrate a technique to generate a flat-top light profile inside a bow-tie cavity, composed of two curved mirrors and two flat mirrors. The flat-top profile is realized by the superposition of two resonant modes of the cavity: $HG_{00}$ and $HG_{10}$ ($HG_{01}$). The beat note produced by the interference of the two resonant fields, oscillates in a time much shorter than any typical atomic motion, giving an effective total intensity that is the sum of the intensities of the two original fields. The transverse modes are not degenerate because the ring cavity is not confocal. The present technique achieved a flat-top field in an empty cavity, guaranteeing at the same time a high-quality-factor cavity and an effective intensity that remains invariant under propagation \cite{bhebhe18}.

\section{Ring cavity design}

\label{sec:design}
\begin{figure}
\centering
\includegraphics[width=\linewidth]{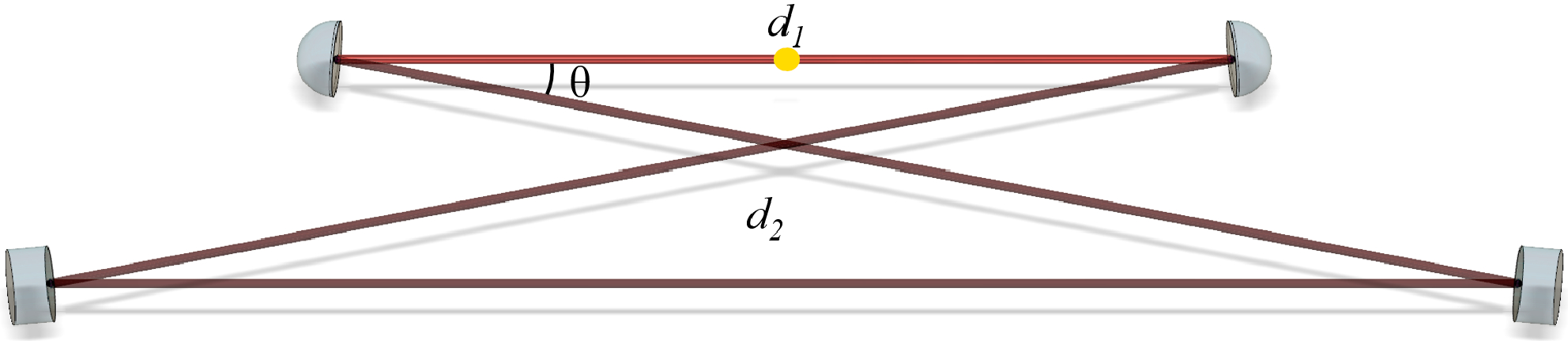}
\caption{Ring cavity. The yellow dot in the middle of $d_1$ corresponds to a hypothetical atomic cloud.}
\label{ringimage}
\end{figure}

The ring cavity is formed by four mirrors, two curved mirrors (radius of curvature $R=50$ cm) separated by a distance $d_1=9$ cm and two flat mirrors giving the remaining cavity length $d_2=39.4$ cm (Fig. \ref{ringimage}). We consider the case of a Gaussian beam propagating in the $z$ direction described by the envelope

\begin{equation}
    A(r)=\frac{A_1}{q(z)} \exp{\left( -ik \frac{\rho^2}{2q(z)} \right)},
    \label{gaussian}
\end{equation}
where $A_1$ a constant, $k=2\pi/ \lambda$, $\rho$ the transverse coordinate, and

\begin{equation}
    \frac{1}{q(z)}= \frac{1}{\mathscr{C}(z)} -i \frac{\lambda}{\pi w^2(z)},
\end{equation}
where $\mathscr{C}$ is the wave front curvature, $\lambda$ the wavelength and $w(z)$ the radius of the transverse mode.

Starting the propagation in the midpoint between the two curved mirrors (corresponding to the position of the atomic trap), it gives an $\mathrm{ABCD}$ matrix for the round trip of \cite{teich, weber}
{\small
\begin{equation}
    M=\begin{bmatrix}
    1+ \frac{1}{f} (\frac{d_1 d_2}{2f} - d_1 - d_2) & d_1 + d_2 + \frac{1}{f} (\frac{d_1^2 d_2}{4f} - \frac{d_1^2}{2} - d_1 d_2) \\
    \frac{1}{f} (\frac{d_2}{f} -2) & 1+\frac{1}{f} (\frac{d_1 d_2}{2f} - d_1 - d_2)
    \end{bmatrix}
    \label{matrizABCD}
\end{equation}}
\hfill \break with $f$ the focal length of the curved mirrors. We find the stationary mode by having the same $q(z)$ value after a round trip (with $\mathrm{A}$, $\mathrm{B}$, $\mathrm{C}$ and $\mathrm{D}$ the elements of the matrix in Eq. \ref{matrizABCD})

\begin{equation}
    q_2 = \frac{\mathrm{A} q_1 + \mathrm{B}}{\mathrm{C} q_1 + \mathrm{D}} = q_1.
    \label{roundq}
\end{equation}
Due to the cavity symmetry, the minimum waist is obtained in the middle of the paths $d_1$ and $d_2$. This gives a collimated beam at the position of the atomic trap, and therefore, an axially homogeneous intensity.

The ring cavity has astigmatic aberrations due to the angle $\theta$ (Fig. \ref{ringimage}). We need to separate the calculation for the tangential ($T$) and sagittal ($S$) planes. The difference for both is in the effective radius of curvature of the curved mirrors \cite{white}

\begin{eqnarray}
    \frac{2 f_T}{R} = \cos (\theta/2) = c & & \frac{2 f_S}{R} = \frac{1}{\cos (\theta/2)} = \frac{1}{c}.
    \label{curvatureR}
\end{eqnarray}
We need to fulfill the stability condition $|A+D| \leq 2$ \cite{skettrup05, teich, weber} of the cavity for both planes, that is

{\small
\begin{align}
    0 \leq \left( 1- \frac{d_1}{cR} \right) \left( 1- \frac{d_2}{cR} \right) \leq 1 &&  0 \leq \left( 1- \frac{c d_1}{R} \right) \left( 1- \frac{c d_2}{R} \right) \leq 1.
    \label{stabilityring}
\end{align}}

\begin{figure}[ht]
\centering
\includegraphics[width=\linewidth]{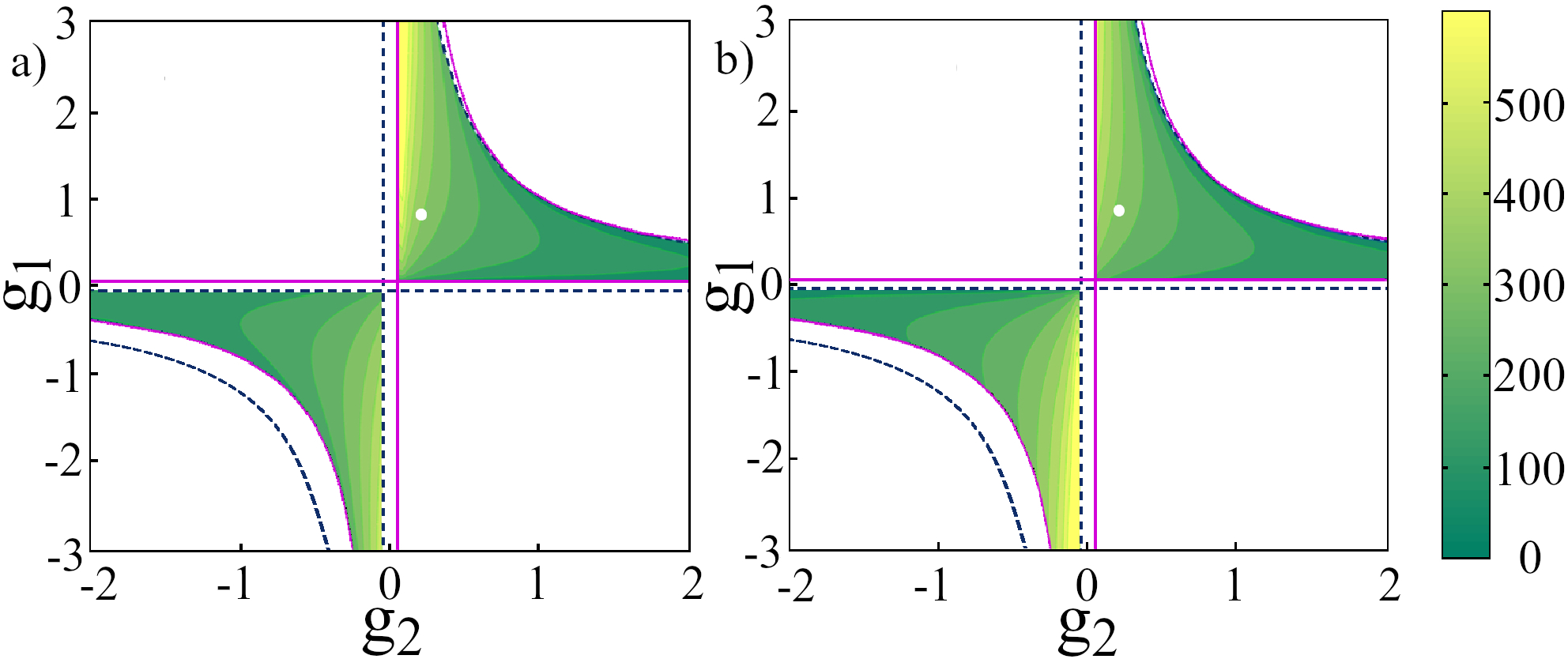}
\caption{Stability region for the ring cavity with $\theta = 15^{\circ}$. The solid pink (dashed blue) line corresponds to the tangential (sagittal) stability boundaries (Eq. \ref{stabilityring}). The filled zone corresponds to the stable region with the colors corresponding to the waist size in $\mu$m (Eq. \ref{cinturas}) for a) the tangential and b) sagittal planes. The white dot corresponds to the cavity in this work.}
\label{regionstability}
\end{figure}

The stability region corresponds to the colored zone in Fig. \ref{regionstability} where we have defined $g_1=1 - d_1/R$ and $g_2=1 - d_2/R$. The waist at the position of the atoms for the tangential and sagittal planes is given by \cite{skettrup05}
{\small
\begin{align}
    w_T^4 = \frac{w_c^4 x (c-x)(c+ \alpha c - \alpha x)}{c- \alpha x} {\rm ;} \hspace{.5em} w_S^4 = \frac{w_c^4 x (1-cx)(1+ \alpha - \alpha cx)}{c(1- \alpha cx)},
    \label{cinturas}
\end{align}}
\hfill \break with $w_c^2= \lambda R/2 \pi$, $x=d_1/R$ and $\alpha = d_2/d_1$.  In our case we obtain a value of $w_T = 349$ $\mu$m and $w_S = 345$ $\mu$m (white point in Fig. \ref{regionstability}). The color in Fig. \ref{regionstability} shows the waist size as a function of the design parameters.

\begin{figure}[ht]
\centering
\includegraphics[width=\linewidth]{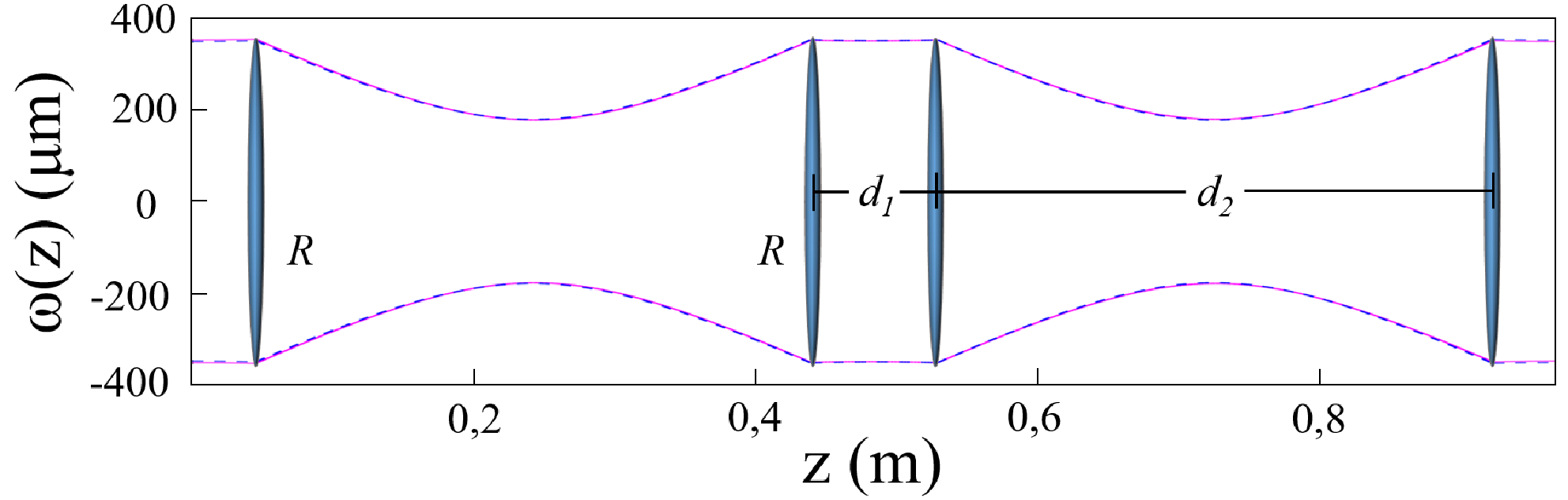}
\caption{Beam waist as it propagates through the unfolded cavity with the curved mirrors represented by lenses. The solid pink (dashed blue) curve corresponds to the tangential (sagittal) plane.}
\label{cavityunfolded}
\end{figure}

Fig. \ref{cavityunfolded} shows the beam waist as it propagates through the unfolded cavity. $z=0$ corresponds to the middle point between the curved mirrors (see Fig. \ref{ringimage}), which is where the atoms are trapped. $d_1+d_2$ corresponds to a round trip. Each mirror is replaced in this drawing by a lens of the corresponding focal length ($f_T=Rc/2$ and $f_S=R/2c$). We want a beam as collimated as possible in $d_1$ to have an axially constant intensity, and a waist ($w_T$ and $w_S$) big enough to cover the atomic cloud that has a diameter of approximately 300 $\mu m$. As can be seen from Fig. \ref{cavityunfolded}, a big collimated beam in $d_1$ implies that $d_2 \simeq 2f$. In contrast to a Fabry Perot cavity, here we can play with the two distances ($d_1$ and $d_2$) to adjust the beam waist and have it collimated. The waist size in the tangential plane diverges as $g_2$ approaches its corresponding stability boundary (vertical solid pink line in Fig. \ref{regionstability}). The waist size in the sagittal plane does not diverge but only reaches a maximum since the corresponding boundary for that plane (vertical dashed blue line in Fig. \ref{regionstability}) lies outside the stability region. The maximum waist in this plane for the cavity angle and $d_1$ value we use is 446 $\mu$m and depends weakly on $g_1$. The increase in sensitivity to alignment and vibrations near the stability boundary is not too big for the cavity configuration we are using \cite{carstens13}. Other solution to have a big waist includes adding a lens inside of a Fabry Perot cavity \cite{riou17}. Using the ring cavity allows reaching a high-quality-factor, something important if one needs to have a high field and a very clean mode. A ring cavity also allows having frequency chirps in the input beams, as opposed to the fixed boundary conditions in a Fabry Perot cavity.

Increasing $\theta$ angle makes the ring cavity more astigmatic and changes the separation of the transverse modes. The value of $\theta=15^{\circ}$ that we use gives a quasi-uniform waist for the two planes (Fig. \ref{cavityunfolded}). There are additional geometrical constraints imposed by having the mirrors out of the way of all the other beams (trap, optical pumping, Raman transition, imaging, etc.) needed for experiments with trapped atoms.

\section{Flat-top profile synthesis}

We obtain an approximation to a flat-top transverse profile by combining two transverse modes of the cavity. We excite the Hermite-Gauss ($HG_{mn}$) modes of the cavity that have an electric field given by \cite{naik18,teich}.

{\small
\begin{eqnarray}
    E_{mn}(x,y,z)=E_0 \frac{w_0}{w(z)} H_m \left( \frac{\sqrt{2} x}{w(z)} \right) H_n \left( \frac{\sqrt{2} y}{w(z)} \right) \exp \left( -\frac{x^2+y^2}{w(z)^2} \right) \nonumber \\
     \exp \left( -i \left[ kz - (1+n+m) \arctan \left( \frac{z}{z_R} \right) + \frac{k (x^2 +y^2)}{2\mathscr{C}(z)} \right] \right),
    \label{HGmodes}
\end{eqnarray}}
\hfill \break with $H_n$ the Hermite polynomials and $z_R$ the Rayleigh length. The astigmatism of the cavity breaks the cylindrical symmetry and gives rise to Hermite-Gauss functions in the sagittal and tangential directions. We combine the modes $HG_{00}$ and $HG_{10}$ ($HG_{01}$) in the correct proportion to approximate a flat-top profile in the horizontal (vertical) direction.

We couple to higher-order Hermite-Gauss beams by starting from a Laguerre-Gauss beam $LG_{10}$ that is generated by sending a Gaussian beam through a single mode fiber with a spiral phase plate attached at the output (aBeam technologies), designed to produce a topological charge of 1 at 780 nm. This beam is a 50-50 superposition of the $HG_{01}$ and $HG_{10}$ modes, so that the same beam can be used to get a flat-top in the vertical or horizontal directions. We couple the $HG_{00}$ and the $HG_{10}$ to the cavity and we control the amount of each transverse mode independently. Both beams come from the same laser, but the $HG_{00}$ has been shifted in frequency with respect to the $HG_{10}$ to have them both resonant with the cavity at the same time. This frequency difference of 230.00(4) MHz gives an interference between the two modes that oscillates in time at a rate much faster than the typical atomic motion. The atoms see an effective field with the interference term averaged out, corresponding to the sum of the intensities of the modes $I_{00}$ and $I_{10}$ with weights given by constants $a$ and $b$. This gives a robust method since there is no need for interferometric control of the paths of the two transverse modes. The total intensity at the waist is given by
\begin{equation}
    I = aI_{00} + bI_{10} = I_0 \left(a + b \frac{4 x^2}{w_0^2} \right) \exp \left( -\frac{2(x^2+y^2)}{w_0^2} \right),
    \label{HG0010}
\end{equation}
with $I_0$ a constant. We obtain a flat-top profile by making the second derivative at the origin equal to zero, that is

\begin{equation}
    \frac{d^2I}{dx^2}|_{x=0} = 0 = a - 2 b.
    \label{conditionflat}
\end{equation}

\section{Experimental setup and results}

\begin{figure}[ht]
\centering
\includegraphics[width=\linewidth]{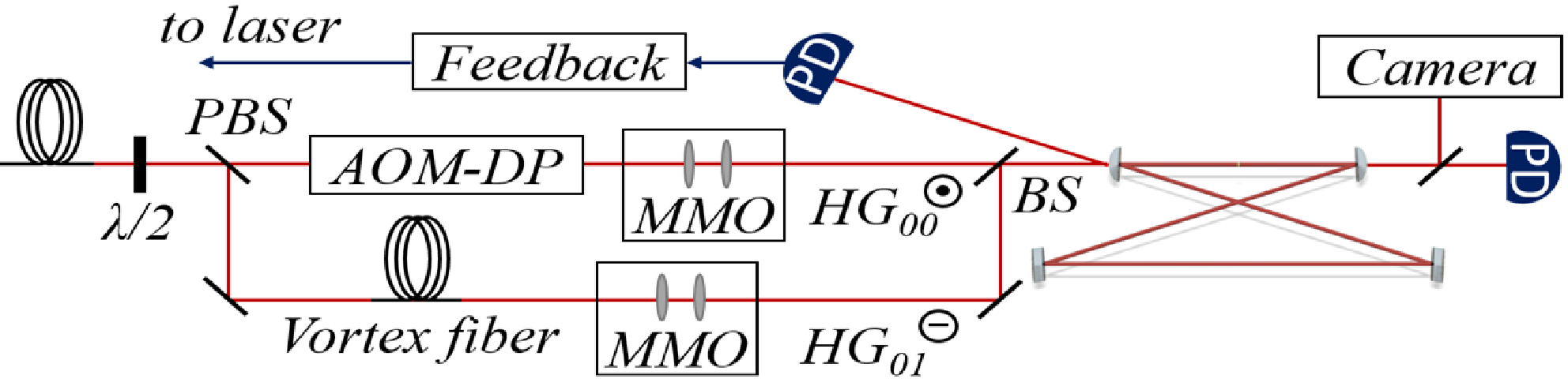}
\caption{Experimental setup for the flat-top profile implementation. BS: beam splitter, PBS: polarizing beam splitter, AOM-DP: acousto-optical modulator in double-pass configuration, MMO: mode matching optics PD: photodiode, and the polarization of the beams before entering the cavity is shown.}
\label{flattopsetup}
\end{figure}

Figure \ref{flattopsetup} shows the optical setup used for the generation of the flat-top profile. We couple light at 860 nm from a Ti:Saph laser (SolsTiS-3100-SRX-F) through a polarization maintaining single mode fiber and we split it into two beams. The light for the $HG_{00}$ beam goes through a double pass acousto-optic modulator to shift the frequency of that beam by 230.00(4) MHz. We send this beam (with $M^{2}=1.01(7)$) through lenses to mode match the beam to the ring cavity. The other path is coupled to the fiber (with polarization control) with the spiral plate at the end to produce the $LG_{10}$ mode. We also mode match this beam (with $M^{2}=0.98(8)$) before we send it to the cavity. Both beams have perpendicular polarizations, but other polarization combinations can be used since the interference term of both beams does not survive as we have already discussed. We monitor the transmitted light with a photodiode and a camera. The reflectivity of all the cavity mirrors is the same, and this gives an imperfect impedance matching and a theoretical transmission on resonance of 75\%.

\begin{figure}[ht]
\centering
\includegraphics[width=\linewidth]{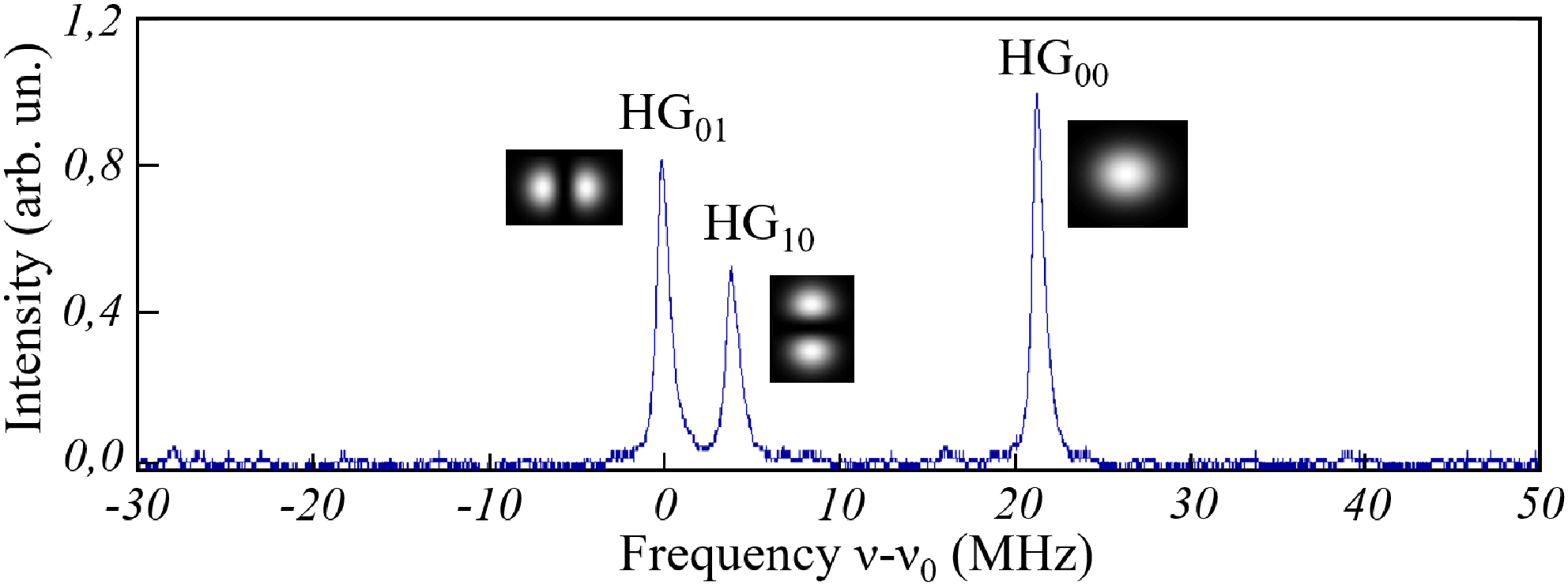}
\caption{Frequency scan of the cavity. The peaks for the $HG_{00}$, $HG_{10}$ and $HG_{01}$ are indicated, and they repeat every Free Spectral Range of 620(2) MHz.}
\label{scanfreq}
\end{figure}

The cavity is isolated mechanically and thermally from the optical table by alternating layers of Sorbothane and aluminum and it is enclosed in a housing of acoustic foam insulation. This isolation strongly reduces cavity fluctuations. Figure \ref{scanfreq} shows the spectrum of the transmitted light as we scan the frequency of the laser. Here we set a frequency difference of 207.95(4) MHz between the two beams that go into the cavity with the acousto-optic modulator. At this frequency, the peaks corresponding to the modes $HG_{00}$, $HG_{10}$ and $HG_{01}$ are clearly resolved. There is a separation between the modes $HG_{10}$ and $HG_{01}$ of 3.97(7) MHz due to the cavity astigmatism. The mode matching suppresses most of the other transverse modes. We overlap the $HG_{00}$ and $HG_{10}$ modes by setting their frequency difference back to 230.00(4) MHz to approach a flat-top in the horizontal direction. We can do the same on the vertical direction by overlapping the $HG_{00}$ and $HG_{01}$ modes with a frequency difference of 226.03(4) MHz.

\begin{figure*}[ht]
\centering
\includegraphics[width=1\textwidth]{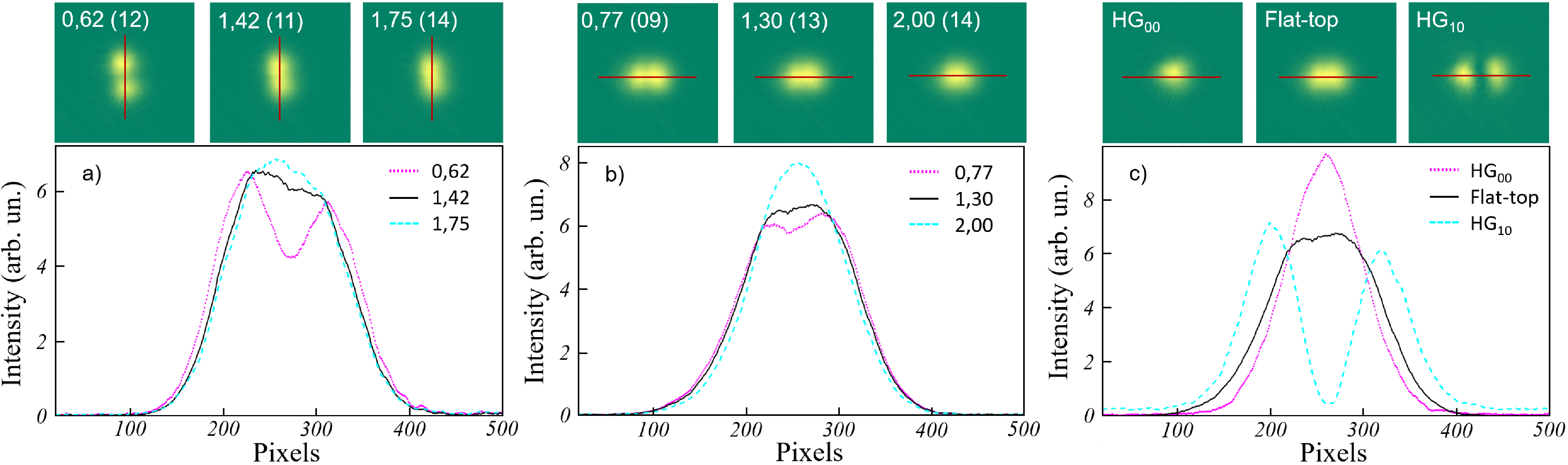}
\caption{Flat-top synthesis in the a) vertical and b) horizontal directions. The acousto-optical modulator was tuned to 226.03(4) MHz and 230.00(4) MHz respectively. The power ratio between the  $HG_{00}$ and the $HG_{10}$ (or $HG_{01}$) modes is indicated in each figure (with the uncertainty indicated in parenthesis). c) Comparison of the $HG_{00}$, $HG_{10}$ modes and flat-top.}
\label{resultflattop}
\end{figure*}

We lock the laser to the cavity to have a stable field. To do this we modulate the frequency of the acousto-optic modulator on the $HG_{00}$ path at a rate of 150 kHz and with an amplitude of 200 kHz. We monitor the light reflected from the first cavity mirror with a photodiode and we send that signal to the locking electronics \cite{arias18}. We use the Pound-Drever-Hall technique \cite{drever83} to generate the feedback signal that we send back to the laser. Since the modulation only affects the $HG_{00}$ beam, it is this beam the one that is locked to the cavity.

Figure \ref{resultflattop} shows the image of the transmitted field through the cavity. As we change the power ratio of the $HG_{00}$ to the $HG_{10}$ modes, we show that we approach a flat-top profile both in the vertical and horizontal directions (Figs. \ref{resultflattop} a and b). The figure shows the images as well as the profile along the line indicated (averaging over 30 pixels in the perpendicular direction). We determine the power ratio of the two modes by comparing the height of their resonances in the transmitted spectrum. There is an imbalance in the two lobes intensity over the $HG_{10}$ and $HG_{01}$ images, probably due to the camera response or fringing effects from the optical components. This imbalance leads to an image of a non-horizontal flat-top, even when that would not be the case inside of the ring cavity. The $HG_{00}$ mode has a spectral width about two times bigger than the $HG_{10}$ mode. This difference might explain why we get a flat-top at an apparent power ratio between the two modes smaller than 2. We give a comparison between the $HG_{00}$, $HG_{10}$ modes and flat-top (Fig. \ref{resultflattop} c) to show the improvement in the flat region. The profile is flat to 1.0(2) \% and 0.6(2) \% for the vertical and horizontal directions respectively over a region of 0.85 times the Gaussian width $\omega_{S}$, an improvement compared to other methods \cite{liang09}.\vspace{-4pt}

\section{Conclusions}
We demonstrate a method to approach a flat-top profile inside a ring cavity by combining two transverse modes. The technique gives a very homogeneous field as big as the size of a typical atomic trap. This method allows for a high-quality-factor and a very clean mode inside of the cavity without the need for interferometric control of the two modes. Our  technique  can  be  extended  to  a  flat-top  with  cylindrical  symmetry  for cavities that support Laguerre-Gauss modes. This design opens the possibility to have a standing wave whose location can be modified by changing the phase of the input beams.\\ \hfill \break
\textbf{Funding}. Consejo Nacional de Ciencia y Tecnología (CONACyT) (157, 225019, 254460, 260704); Universidad Autónoma de San Luis Potosí (UASLP) (FAI-UASLP).\\ 
\textbf{Acknowledgment}. The authors thank Israel García, Diego Lancheros and Javier González for their assistance.

\bibliography{references}
\bibliographyfullrefs{references}

\end{document}